\documentclass[pre,twocolumn,final, showpacs, floatfix]{revtex4}
\usepackage{graphicx}
\usepackage{amsmath,amssymb,bm}
\usepackage{color}

\usepackage{mathptmx}

\usepackage{psfrag}
\usepackage{verbatim}

\newcommand{\expt}[1]{\langle #1 \rangle}
\newcommand{\Expt}[1]{\left\langle #1 \right\rangle}

\newcommand{\beq}{\begin{equation}}
\newcommand{\eeq}{\end{equation}}

\newcommand{\bp}{\mathbf{p}}

\begin{document}

\title{Bottom-up derivation of an effective thermostat for united atoms simulations of water}
\author{Anders Eriksson}
\affiliation{Department of Physics, University of Gothenburg, 41296 G\"oteborg, Sweden} 
\author{Martin Nilsson Jacobi}
\author{Johan Nystr\"{o}m}
\author{Kolbj\o rn Tunstr\o m}
\affiliation{Complex Systems Group, Department of  Energy and Environment, Chalmers University of Technology, 41296 G\"oteborg, Sweden} 


\pacs{47.11.+j, 05.40.-a, 45.70.-n, 47.10.+g}


\begin{abstract}
In this article we derive the effective pairwise interactions in a Langevin type united atoms model of water. The interactions are determined from the trajectories of a detailed molecular dynamics simulation of simple point charge water. A standard method is used for estimating the conservative interaction, whereas a new ``bottom-up'' method is used to determine the effective dissipative and stochastic  interactions. We demonstrate that, when compared to the standard united atoms model, the transport properties of the coarse-grained model is significantly improved by the introduction of the derived dissipative and stochastic interactions. The results are compared to a previous study, where a ``top-down'' approach was used to obtain transport properties consistent with those of the simple point charge water model.
\end{abstract}

\maketitle

\section{Introduction}

Water is the most important biological solvent in natural systems. Accurate molecular dynamics modeling of, for example proteins or membranes, depend critically on how well the model captures the properties of the ambient water. Unfortunately, the structure of the water molecule, especially its tendency to build local hydrogen networks, makes it hard to accurately model liquid water. A large number of different molecular dynamics (MD) models have been proposed, each with advantages and shortcomings, and often tailored to fit a particular use. See e.g.\citep{Guillot02,jorgensen05,water_web}  for an overview. In addition to these difficulties, all atomistic models of water share the problem of being too computationally demanding for modeling the time and length scales that are relevant for many self-assembly and folding processes in biochemistry. To circumvent these limitations several coarse-grained models of water have been suggested in the literature, for example the united atoms model that we are considering in this paper. While it is clear that many central properties of water can only be modeled implicitly in the aggregated models, especially properties depending on the extended hydrogen networks, the models' computational efficiency enables the study of processes which occur on scales unattainable using more detailed models.

In this article we consider how to determine the effective deterministic and stochastic interaction potentials in a united atoms representation of water. Generically, a coarse-grained model of a mechanical system is described by Langevin dynamics. Under certain conditions, for example when the degrees of freedom removed by the coarse-graining are averaging, the resulting dynamics is deterministic, but in general both noise and dissipation must be included in the reduced representation. It is somewhat surprising that this fact is often ignored when effective potentials for united atoms models of molecular systems are derived. Under the assumption that the effective interactions are pairwise, an important result of Henderson guarantees a one-to-one correspondence between the pair correlation function and the pairwise potential \citep{Henderson74}. Using this result, measurements of center of mass pair correlations from detailed molecular dynamics simulations can be used to determine the conservative potential to be used in a united atoms simulation \citep{lyubartsev03, lyubartsev1995, ilpo04, soper96, reith_etal03, almarza_lomba03} (alternatively, direct time averaging over the fast degrees of freedom can be used, see e.g. Refs. \onlinecite{forrest_suter95,izvekov_parrinello04}). Importantly, the equilibrium pair distribution is not altered by the noise and dissipative interactions \citep{eriksson09}. This is perhaps the reason why these interactions are often ignored or represented by a weak general purpose thermostat.

The elegant result by Henderson also causes problems by the simple fact that it is completely general. Since the pair correlation function can usually be measured in a detailed simulation, we can also usually derive an effective pairwise potential, even if the coarse-grained representation is not closed and therefore has no meaning. The conclusion is that requiring the coarse-grained interactions to recreate the correct pair distribution does not guarantee any correspondence between the dynamic properties of the detailed system and the coarse-grained system. We may even argue that it is not clear if it is worth while to simulate the exact time dynamics of the coarse-grained model; If the correspondence between the detailed and the coarse-grained system is only their respective equilibrium configurations, it is probably more efficient to use a Monte Carlo algorithm to minimize the free energy and to sample from the ensemble of equilibrium configurations.

We have demonstrated in earlier work that the transport properties of simple point charge water (SPC), simulated with MD, and the corresponding united atoms model, based on Henderson's theorem, differ significantly \citep{eriksson08b}. In the same study it is was shown that by adding a DPD thermostat to the united atoms model, the transport properties can be tuned to correctly represent those of the detailed MD simulation, while at the same time preserving the equilibrium pair correlation function. The motivation for the DPD thermostat is that it is the most general pairwise interaction that respects the local symmetries in a mechanical system, i.e., conservation of momentum and angular momentum. The disadvantage of this approach is clearly that the correspondence between the detailed dynamics and the coarse-grained dynamics may be lost.  

In a recent publication we introduced a method for determining the stochastic and dissipative interactions in systems simulated with a dissipative particle dynamics (DPD) thermostat \citep{eriksson09}.  The method can be viewed as a compliment to the inverse Monte Carlo technique \citep{lyubartsev1995,lyubartsev03} used to reconstruct the conservative interaction based on Henderson's result. Together, these two methods can be used to determine the detailed structure of the effective conservative, dissipative, and  stochastic interactions in a united atoms model. In addition, the framework gives a clear indication on whether or not the united atoms representation is a valid coarse-graining. 

In this paper, we apply the above framework to a united atoms model of SPC. This is possible using the assumption that the reduced degrees of freedom are not averaged, but rather leads to noise and dissipation that can be represented by the stochastic and dissipative interactions in a DPD thermostat. This assumption should be viewed as similar to the pairwise potential assumption in Henderson's theorem. The transport properties of the resulting united atoms-DPD model are much improved compared to the original united atoms model. We discuss the implications of this result for the understanding of united atoms models and how well a DPD thermostat manages to represent the reduced degrees of freedom.

\section{Theoretical background}

\subsection{DPD}

DPD was introduced in 1992 by Hoogerbrugge and Koelman~\cite{hoogerbrugge_koelman92, koelman_hoogerbrugge92} as a particle based simulation technique for complex fluids. It was originally considered as a blend between the Molecular Dynamics and Lattice Gas Automata techniques, the first because of the similar integration scheme and representation, and the latter because it is focused mainly on respecting the mechanical conservation laws.

The DPD equations of motion, with particle positions $\mathbf{r}_{i}$, velocities  $\mathbf{v}_{i}$ and momenta $\mathbf{p}_{i}$, can be written as a system of Langevin equations
\begin{align}
	\label{eq: rdot DPD}
	\dot{ \mathbf{r} }_{i}    &= \mathbf{v}_{i}, \\
	\label{eq: pdot DPD}
	\dot{ \mathbf{ p } }_{i}  &= \sum_{j \neq i} \left[ \mathbf{F}_{ij}^\text{C} + \mathbf{F}_{ij}^\text{D}
										  +\mathbf{F}_{ij}^\text{S} \right],
\end{align}
where $\mathbf{F}_{ij}^\text{C}$, $\mathbf{F}_{ij}^\text{D}$ and  $\mathbf{F}_{ij}^\text{S}$ are conservative, dissipative, and stochastic forces between particles $i$ and $j$. A particular property of DPD is that all interactions are modeled by pairwise central forces. This ensures the conservation of linear and angular momentum which in turn implies the correct Navier-Stokes equations in the continuum limit \cite{espanol95, marsh_etal97_FPB}.

In the original formulation of DPD, the dissipative and stochastic forces were chosen independently. In what later has become the standard DPD model, formulated by \citet{espanol_warren95}, the dissipative and stochastic forces are not independent; rather, they depend on each other through a fluctuation dissipation theorem and together function as a  local pairwise thermostat. The equilibrium properties of a DPD system are determined by the conservative force alone, while the thermostat defines how the system reaches equilibrium. The common way to  express the dissipative and stochastic forces is
\begin{align}
	\label{eq: Dissipative force}
	\mathbf{F}_{ij}^\text{D} &= 
	- \omega^2(r_{ij}) \, \left(\mathbf{e}_{ij} \!\cdot\!\mathbf{v}_{ij}\right)\, \mathbf{e}_{ij}, \\
	\label{eq: Stochastic force}
	\mathbf{F}_{ij}^\text{S} &=  \sqrt{2k_B T} \omega( r_{ij} ) \, \zeta_{ij} \, \mathbf{e}_{ij},
\end{align}
where $r_{ij}$ is the distance between particles $i$ and $j$, $\mathbf{e}_{ij}$ is the unit vector pointing from $j$ to $i$, and $\mathbf{v}_{ij}$ is the velocity difference $\mathbf{v}_{ij} = \mathbf{v}_j - \mathbf{v}_i$. $k_\text{B}$ is Boltzmann's constant, and $T$ is the temperature. The scalar function $\omega(r_{ij})$ describes how the stochastic and dissipative forces depend on the distance between the particles, and $\zeta_{ij}$ is interpreted as a symmetric Gaussian white noise term with mean zero and covariance
\begin{equation}\label{eq:zeta_cov}
	\expt{\zeta_{ij}(t)\zeta_{i'j'}(t')} = (\delta_{ii'}\delta_{jj'} + \delta_{ij'}\delta_{ji'})\delta(t-t'), 
\end{equation}
where $\delta_{ij}$ and $\delta(t)$ are the Kronecker and Dirac delta functions, respectively. 

It has become standard practice to choose the conservative and stochastic interactions in DPD to be linearly decreasing functions with a finite cutoff radius. To give the simulated system a physical interpretation the magnitude of the conservative force is generally tuned such that some desired equilibrium property of the system, e.g. the compressibility \cite{groot_warren97}, is matched. The magnitude of the stochastic force is then set such that the temperature equilibration is reasonably fast, but still slow enough to allow a stable simulation.

We and others have previously argued that the DPD ansatz is to be seen as the result of a systematic coarse-graining from a given microscopic system and that neither the conservative nor the dissipative/stochastic interactions can be chosen arbitrarily~\cite{flekkoy_etal99,espanol03,eriksson09}. Rather, the DPD interactions should result from projecting the microscopic dynamics onto a coarse-grained level. This argument is valid regardless of the projection, as long as the projected dynamics truly follows the DPD ansatz. Types of projections that can be considered are for example clustering collections of freely moving particles into single DPD beads, which is the way DPD is commonly viewed~\cite{hoogerbrugge_koelman92, espanol_warren95, travis_etal08, groot_warren97, allen06, peters04}, or DPD used as a thermostat for united atoms coarse-graining~\cite{lyubartsev03, praprotnik_etal07a, soddemann_etal03}. In the former case, the stochastic and dissipative interactions result from the internal motion of the microscopic particles inside the DPD beads, whereas for the latter case, these interactions result from the internal atomistic motion not explicitly simulated in the coarse-grained particles.

\subsection{Deriving the DPD forces}

The effective interactions in the DPD model is derived using the trajectories from a detailed molecular dynamics simulation. The trajectories of the particles in the united atoms representation are calculated as a center of mass projection from the atomistic trajectories. The goal of our method is to find forces acting on the center of mass point particles that give equilibrium distribution and transport properties that are as close as possible to those caused by the true force field.

We use a tiered strategy, where we first find the radial distribution function (RDF) of the united atoms system, calculated from the simulations of the underlying SPC system. 
Given this function, we use the Inverse Monte Carlo (IMC) method \citep{lyubartsev1995,lyubartsev03} to find the potential function that gives rise to the same RDF. This and similar methods rest on the result by \citet{Henderson74}, who showed the uniqueness of such a potential. The IMC method is an iterative method which may be viewed as a multi-dimensional Newton-Raphson technique. In each iteration the potential is updated using the gradient of the RDF with respect to the potential, which is estimated from standard forward Monte Carlo simulations using the present potential.

Once the IMC method has converged to a solution, the conservative force is fixed. In order to complete the DPD description of the united atoms system, it remains to determine the dissipative and stochastic forces. Given that these two forces obey a dissipation-fluctuation relation, the choice of stochastic force does not influence the equilibrium distribution, only the approach to equilibrium and hence, in general, the transport properties of the system. Unfortunately, this also means that the RDF gives us no information about the relevant stochastic and dissipative forces, and we must turn to observables that more directly reflects how the shape and magnitude of the stochastic force determines the transport properties. 

In \citep{eriksson09} we have suggested a practical scheme based on the auto-correlation of the force fluctuations, $\kappa_F(r,\tau)$ , defined as
\begin{equation}
\label{eq: kappa_F}
	\kappa_F(r,\tau) = \Expt{\delta_t (\delta_F \bp_i)\cdot\delta_t (\delta_F \bp_j)}_{r},
\end{equation}
where $\expt{}_{r}$ denotes ensemble average conditional on the distance $r$ between particles $i$ and $j$ at time zero, and where
\begin{equation}
\label{eq: delta_t delta_F p_i}
	\delta_t (\delta_F {\bf p}_i) = \bp_i(t) - \bp_i(0) - \int_0^{\tau} \!dt \left[{\bf F}_i^C(t) + {\bf F}_i^D(t)\right] 
\end{equation}
is the contribution to particle $i$'s change in momentum from the stochastic forces during the time interval $[0,\tau]$. [In our notation $\delta_t$ denotes a difference in time, and $\delta_F$ the difference between total force and deterministic (conservative and dissipative) forces.] 

Consider now the thought experiment of adding a force to the SPC simulation to freeze the center of mass of each water molecule in place at time zero, but allowing the internal degrees of freedom of each molecule to evolve as usual (i.e. the orientation and internal vibration of the molecules). Using Mori-Zwanzig theory one can then show the exact relation \citep{eriksson09}
\begin{equation}
\label{eq: consistency equation}
	\omega^2(r) = \lim_{\tau \rightarrow \infty}-\frac{1}{2 k_B T} \frac{\partial}{\partial \tau} \kappa_F(r,\tau)
\end{equation}
where $\omega(r)$ is the function determining both the dissipative and the stochastic DPD interactions. In the SPC simulations, no such force is added, but Eq.~(\ref{eq: consistency equation}) may still hold approximately in a region of $\tau$ if the internal degrees of freedom change on a faster time scale than the center of mass positions of the molecules, i.e., if there is a region where $\kappa_F(r,\tau)$ is an approximately linear function of time. In \citep{eriksson09} we showed that this is indeed the case if the DPD ansatz holds. In the underlying SPC system, the DPD assumption that the time autocorrelation of the center of mass forces is negligible does not hold for short time scales. In this case, the time region must be chosen sufficiently long that the force has time to decorrelate, but still short enough that the particles do not move too far from their positions at the beginning of the time interval.

Since the right-hand side of Eq.~(\ref{eq: consistency equation}) depends on the function $\omega(r)$, our method is iterative. As a starting point, we take  $\omega(r) = 0$, corresponding to a strictly deterministic time evolution of the center of masses.  The method is then as follows: First, for each value of $r$ evaluate $\kappa_F(r,\tau)$ as a function of time. Second, identify the time region where $\kappa_F(r,\tau)$ is approximately linear (this region  may be different for different values of $r$), and estimate $\omega(r)$ as the right-hand side of Eq.~(\ref{eq: consistency equation}). If the new and the old $\omega(r)$ are close for all values of $r$,  Eq.~(\ref{eq: consistency equation}) holds approximately and we may stop. Otherwise, we repeat the procedure from step one.
This concludes our background section. Before we present our results, in the next section we give the details of the molecular dynamics simulation of the SPC model.

\section{Simulation setup}
We use Simple Point Charge (SPC) water as the model system. The SPC model is not the most accurate water model available when it comes to representing transport properties~\cite{spoel_etal}, but is widely used and serves as a good test case for our method. The simulations were performed with a periodic cubic box of size $3.0$ nm, with a total of $895$ SPC molecules, using the molecular dynamics software GROMACS. The integration time step was set to $2.0$ fs, and temperature and pressure equilibration was performed using the Berendsen thermostat, with target temperature and pressure $298$ K and $1$ bar respectively. The compressibility parameter was set to $4.6\times 10^{-5}$ (1/bar). Long range electrostatic forces was handled using the reaction-field method, with a cut-off of 1.4 nm. From an initial simulation, the radial distribution function (RDF) of the center of mass of the SPC molecules was measured. This data was then used as input to the inverse Monte Carlo method, which produces an effective potential for the center of mass motion of the coarse-grained representation of the SPC system. Given this potential the GROMACS SPC simulation was reran, now measuring $\kappa_F$ [cf. Eq. (\ref{eq: delta_t delta_F p_i})], and gradually changing $\omega(r)$ until convergence.

\section{Results}
\label{sec: Results}
\subsection{Conservative potential}

\begin{figure} 
\psfrag{r}[][]{$r$ [nm]}
\psfrag{potential}[][]{$\Phi(r)$ [$k_B T$]}
\includegraphics[width=252pt]{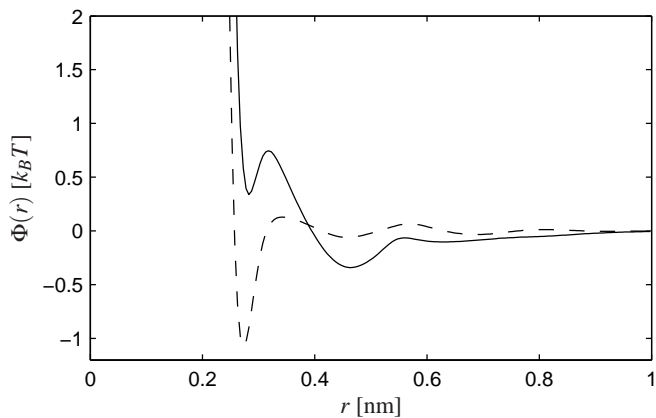}
\caption{\label{fig: conservativePotential}
Pairwise potential derived from the radial distribution function for the center of mass of the SPC molecules by using the IMC method (solid). The potential converges to zero at about $r = 1.0$ nm. The potential of mean force is plotted as a reference (dashed line), deviating substantially from the estimated potential.
}
\end{figure}

Following the procedure in \cite{lyubartsev1995}, we took the potential of mean force, $\Phi_{MF} = -k_B T \log[g(r)]$, as the initial guess for the effective potential, shown as the dashed graph in Fig.~\ref{fig: conservativePotential}. The final effective potential obtained using IMC, is shown as the solid graph in the same figure. This is the result after $10$ iterations, each using $5$ million Monte Carlo updates. The resulting potential is clearly different from the potential of mean force, and has the important property that it faithfully represents the equilibrium properties of the system according to the theorem of \citet{Henderson74}. This is true in any simulation, independent of the thermostat, as long as the thermostat is constructed according to a fluctuation-dissipation theorem~\cite{eriksson09}.

\subsection{Dissipative/Stochastic forces}
The potential derived in the previous section determines the functional form of the conservative forces, $F^C_{ij}$, between the particles. In a simulation, the effective conservative force on a coarse-grained particle can then be calculated from $F^C_i = \sum_{j\neq i}{F^C_{ij}(r_{ij})}$. From Eq. (\ref{eq: delta_t delta_F p_i}) we get an estimate of $\delta_t (\delta_F {\bf p}_i)$ under the initial assumption that the dissipative forces are all equal to zero. By using Eqs.~(\ref{eq: consistency equation}) and (\ref{eq: kappa_F}), a new estimate of $\omega(r)$ can be found. Inserting this new approximation in the dissipative force in Eq. (\ref{eq: delta_t delta_F p_i}), and repeating the procedure until convergence produces the $\omega(r)$ function that should be used together with the conservative potential in a DPD representation of the original SPC system.

\begin{figure} 
\psfrag{t}[][]{$\tau$ [ps]}
\psfrag{DdpiDdpj}[][]{$\kappa_F$ [$10^{-45}$\text{ }kg$^2$ m$^2$/s$^{2}$]}
\includegraphics[width=252pt]{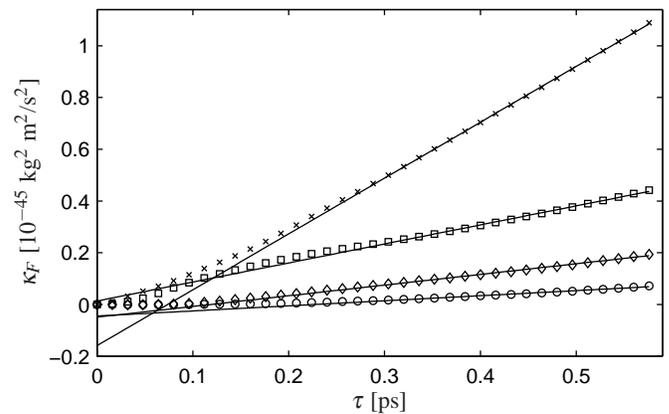}
\caption{\label{fig: DdpiDdpj}
The asymptotic slope of $\kappa_F$ defines, through Eq. (\ref{eq: consistency equation}), the functional form of $\omega(r)$. In this figure, the values of $\kappa_F$ are shown as a function of the time difference, $\tau$ [cf. Eq. \ref{eq: delta_t delta_F p_i})], for four different distance values: $r = 0.27$ nm ($\scriptstyle\times$), $r = 0.33$ nm ($\scriptscriptstyle \square$), $r = 0.39$ nm ($\diamond$),  and $r = 0.45$ nm ($\circ$). The solid lines are linear fits of the $\kappa_F$ function in the range $\tau \in [0.3,\text{ }0.6]$ ps.
}
\end{figure}

Fig.~(\ref{fig: DdpiDdpj}) shows the measured values of $\kappa_F$ for the first iteration step. 
For small times, the motion of the SPC molecules' mass centers are smooth and deterministic as determined by the forces in the detailed SPC model. On this time scale, the white noise approximation of the DPD method is ill-defined. It is only on longer time scales that the degrees of freedom taken away by the projection can be assumed to affect the slow (center of mass) degrees of freedom in the form of noise and dissipation. At these time scales, we expect from theory to find a linear region in $\kappa_F$ \cite{eriksson09}, and estimate $\omega(r)$ from the slope [cf. Eq. (\ref{eq: consistency equation})]. We identify this as the region above $\tau = 0.3$ ps. For times significantly larger than $\tau = 0.6$ ps, the center of mass of the SPC molecules will have had time to move a non-negligible distance from their original positions, rendering the conditioning on $r$ in Eq. (\ref{eq: kappa_F}) ill-defined. The region from which the slope of $\kappa_F$ can be measured must hence be chosen smaller than this time. In our simulations the SPC molecules move on average $0.14$~nm during $0.6$~ps, which is close to half the distance to the first peak in the RDF. Hence, we conclude that the particles stay close to their initial position during the time interval used to determine $\omega(r)$.

\begin{figure} 
\psfrag{omega}[][]{$\omega(r)$ [$\sqrt{(\text{kg}/\text{s})}$]}
\psfrag{r}[][]{$r$ [nm]}
\psfrag{difference from last boot strapping step}[][]{$\Delta \omega_k$}
\psfrag{boot strapping step}[][]{$k$}
\includegraphics[width=252pt]{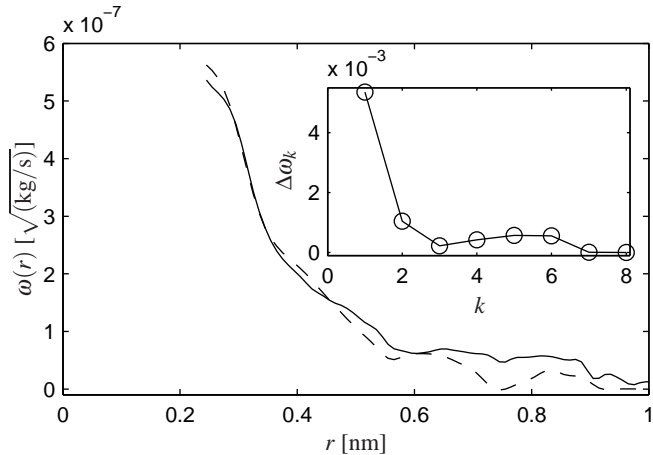}
\caption{\label{fig: omega}
The figure shows $\omega(r)$ as a function of the inter-particle distance $r$, after the first (dashed line) and last (solid line) steps of the iterative procedure. Starting with $\omega(r) = 0$ for all $r$, the slope of $\kappa_F$ is measured (cf. Fig.~\ref{fig: DdpiDdpj}), and a new estimate of $\omega(r)$ is obtained by using Eq. (\ref{eq: consistency equation}). The convergence towards a self consistent solution is shown as inset, with $\Delta \omega_k$ defined in Eq. (\ref{eq: delta omega}). After eight steps, a self consistent solution is reached, resulting in a $\Delta \omega$-value of approximately $0$.
}
\end{figure}

The resulting form of $\omega(r)$ is shown in Fig.~\ref{fig: omega}. The curves represent the first and last steps in the iterative procedure. After eight updates of $\omega(r)$, the procedure has converged to the function shown as the solid line. The inset shows the norm of the distance between $\omega(r)$ of consecutive iterations, defined as
\begin{equation}
\label{eq: delta omega}
	||\Delta \omega_k||  = \sum_{i=0}^N (\omega_k(i)-\omega_{k-1}(i))^2/\sqrt{\sum _{i=0}^N  \omega^2_k(i)  \sum _{i=0}^N  \omega^2_{k-1}(i) } , 
\end{equation}	
where $k$ is the iteration number, and $i$ goes over all points on the $\omega$-curves. 

Using the derived interactions, i.e., the conservative potential and $\omega(r)$, in a  DPD model, we can measure the transport properties of the coarse-grained model. Table \ref{tbl: transport properties} presents the resulting diffusion and viscosity for the DPD model, compared to the original SPC simulation and also to the simple united atoms (UA) model with no dissipative and stochastic forces. We note that both the diffusion and viscosity are off by roughly a factor of four in the united atoms simulation, whereas the addition of the newly derived DPD thermostat reduces that number to $60$ percent for the diffusion and only $20$ percent for the viscosity.

\begin{table}
\begin{tabular}{lcc}
	Method	&	D [$10^{-9}$ $\text{m}^2 \: \text{s}^{-1}$]	&	$\eta$ [cP]	\\
	\hline
	SPC		&	$3.87 \pm 0.04$	&	$0.44 \pm 0.05$ \\
	UA	&	$15.20 \pm 0.14$	&	$0.120 \pm 0.005$ \\
	DPD	& $6.13 \pm 0.07$	&	$0.36 \pm 0.05$ \\
\end{tabular}`

\caption{\label{tbl: transport properties}
This table presents a comparison between the diffusion coefficient and the viscosity for the microscopic system (SPC) and for two coarse-grained representations of the same system (UA and DPD) described in the main text. A clear improvement of the transport properties of the coarse-grained system can be seen when adding the DPD thermostat.
}
\end{table}

\begin{figure}
\psfrag{omega AE units}[][]{$\omega(r)$ [$\sqrt{(\text{kg}/\text{s})}$]}
\psfrag{r [nm]}[][]{$r$ [nm]}
\includegraphics[width=252pt]{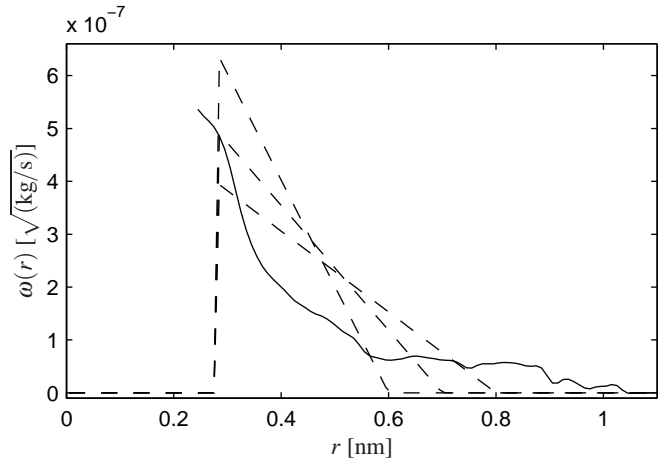}
\caption{\label{fig: compareToTopDown}
The figure shows a comparison between the $\omega(r)$ function derived in this paper and some examples of the functional forms of $\omega(r)$ that were manually tuned to give approximately correct diffusion and viscosity in reference~\cite{eriksson08b}. 
}
\end{figure}

In Fig.~\ref{fig: compareToTopDown} we compare the newly derived $\omega(r)$ function from the bottom-up method presented in this paper with the top-down approach in reference~\cite{eriksson08b}. The dashed curves represent hand tuned $\omega(r)$ functions that all give approximately correct diffusion and viscosity, compared to that of the detailed SPC simulation. The solid line is the $\omega(r)$ function derived without any manual tuning in this paper. The rather close agreement between transport properties of the coarse-grained (DPD) model and the SPC model presented in table \ref{tbl: transport properties} is further strengthened by this figure, as we note that the difference between the derived $\omega(r)$ function and the hand tuned curves is relatively small, though it is worth noting that the derived DPD interactions is consistently weaker than the hand tuned thermostat.  


\section{Discussion}

We conclude with a brief summary and discussion of our results.
First, we have calculated the effective conservative force for a united atoms representation of water from the radial distribution function of the center of mass. Simulations using the resulting forces show that the transport properties of the united atoms model differ significantly from those of SPC water: the diffusion rate is about four times too high, and the viscosity is approximately four times too low. This discrepancy is not too surprising considering that we have replaced the strongly polar SPC water molecule with a simple point particle with a radially symmetric interaction potential.

Second, our iterative method for estimating the stochastic force in the DPD extension of the united atoms model finds the force that best fulfills the self-consistency relation between $\kappa_F$ and $\omega$.
Including the resulting dissipative and stochastic forces gives transport properties that are significantly closer to the SPC simulations. For instance, the DPD diffusion rate is approximately $60\%$ higher than that of the SPC simulations, but excluding the dissipative and stochastic forces yields a diffusion rate that is $290\%$ higher (cf. Table~\ref{tbl: transport properties}). To give some perspective to these figures, consider that the difference between the experimental value of the diffusion rate of water ($2.27 \times 10^{-9}$ m$^2$/s \citep{eisenberg69}) and the diffusion rate of SPC is of the same order as the difference between the diffusion rates of SPC and DPD.

Using a linear ansatz for $\omega(r)$, we showed in a previous report \citep{eriksson08b} that it is possible to obtain a simultaneous agreement of the diffusion rate and viscosity by carefully choosing the support and magnitude of $\omega(r)$. Since these are defined on macroscopic time and space scales, it follows that matching these properties does not necessarily preserve the dynamics at the mesoscale. It is symptomatic that several different parameter settings was found that gave similar transport properties \cite{eriksson08b}. 
The transport properties using the $\omega(r)$ found in this article are not as accurate, but on the other hand no tuning was made of $\omega(r)$ to improve the transport properties. The method presented in this paper is entirely a ``bottom-up'' approach, as opposed to the ``top-down'' method used in stagnated DPD and in \cite{eriksson08b}. Given the constrained nature of the center of mass representation of water, it is not unexpected that we may have to choose which transport property to best represent. The advantage of the method in this paper is that it does not make any assumptions about the shape of $\omega(r)$. If the DPD ansatz is approximately valid, the method automatically identifies the shape, magnitude and support for the stochastic force, without the need to tune any parameters.

The water molecule presents a difficult case for united atoms modeling. Because the dipole-dipole interactions are strong and fluctuate on relatively long time scales (the dipole-dipole decorrelation time is approximately 4.7~ps \citep{Kumar:2006}, much longer than the time over which we estimate the stochastic forces), the center of mass motion of SPC water is neither averaging nor a Langevin process. 
This suggest that we have underestimated the role of the fluctuations. However, the estimated conservative interaction includes information of typical local configurations, and the stochastic interactions capture fluctuations around the effective mean forces. The fluctuations may therefore very well equilibrate around the local mean configurations much faster than the total decorrelation time between the local orientation of the molecules. We think that this is the explanation for the linear region of $\kappa_F$ to appear on relatively short time scales.

Furthermore, we have seen that including the stochastic forces gives a significant improvement over only the conservative forces. This shows that it is possible to find DPD forces that capture not only the equilibrium distribution but also the transport properties of fluids with a rich spatial structure and time evolution, without explicitly representing the polar dynamics. As discussed in the introduction, this can be useful for simulating water in the bulk, where the main concern is the water-water interactions.

\noindent {\bf Acknowledgments}: 
This work was funded (in part) by the EU integrated project FP6-IST-FET PACE, by EMBIO, a European Project in the EU FP6 NEST Initiative, and by the Research Council of Norway. 

\bibliography{coarsegraining}


\end{document}